\documentclass[fleqn,12pt]{article}
\usepackage{amsfonts}
\usepackage{graphics,graphicx} 
\usepackage{amsmath}
\usepackage{amssymb,latexsym}

\usepackage{color}

\usepackage{amsfonts,amssymb,cite}
\usepackage{graphicx}

\def\onecol{\onecolumn \mathindent 2em}

\def\noi{\noindent}

\newcommand{\Title}[1]{\noi {{\Large\bf #1}}\\[1ex]}

\newcommand{\Author}[2]{\noi{\bf #1}\\[2ex]\noi{\normalsize\it #2}\\}

\newcommand{\Abstract}[1]{\vskip 2mm \begin{center}
        \parbox{16.4cm}{\small\noi #1} \end{center}\medskip}
\newcommand{\foom}[1]{\protect\footnotemark[#1]}

\def\email#1#2{\footnotetext[#1]{e-mail: #2}\addtocounter{footnote}{1}}

\def\nqq{\hspace*{-2em}}
\def\nhq{\hspace*{-0.5em}}

\def\cm{\hspace*{1cm}}
\def\inch{\hspace*{1in}}

\def\Jl#1#2{#1 {\bf #2},\ }

\def\ApJ#1 {\Jl{Astroph. J.}{#1}}
\def\CQG#1 {\Jl{Class. Quantum Grav.}{#1}}
\def\DAN#1 {\Jl{Dokl. AN SSSR}{#1}}
\def\GC#1 {\Jl{Grav. Cosmol.}{#1}}
\def\GRG#1 {\Jl{Gen. Rel. Grav.}{#1}}
\def\JETF#1 {\Jl{Zh. Eksp. Teor. Fiz.}{#1}}
\def\JETP#1 {\Jl{Sov. Phys. JETP}{#1}}
\def\JHEP#1 {\Jl{JHEP}{#1}}
\def\JMP#1 {\Jl{J. Math. Phys.}{#1}}
\def\NPB#1 {\Jl{Nucl. Phys. B}{#1}}
\def\NP#1 {\Jl{Nucl. Phys.}{#1}}
\def\PLA#1 {\Jl{Phys. Lett. A}{#1}}
\def\PLB#1 {\Jl{Phys. Lett. B}{#1}}
\def\PRD#1 {\Jl{Phys. Rev. D}{#1}}
\def\PRL#1 {\Jl{Phys. Rev. Lett.}{#1}}

\def\al{&\nhq}
\def\lal{&&\nqq {}}
\def\eq{Eq.\,}
\def\eqs{Eqs.\,}
\def\beq{\begin{equation}}
\def\eeq{\end{equation}}
\def\bear{\begin{eqnarray}}
\def\bearr{\begin{eqnarray} \lal}
\def\ear{\end{eqnarray}}
\def\earn{\nonumber \end{eqnarray}}
\def\nn{\nonumber\\ {}}
\def\nnv{\nonumber\\[5pt] {}}
\def\nnn{\nonumber\\ \lal }
\def\nnnv{\nonumber\\[5pt] \lal }
\def\yy{\\[5pt] {}}
\def\yyy{\\[5pt] \lal }
\def\eql{\al =\al}

\def\dst{\displaystyle}

\def\fracd#1#2{{\dst\frac{#1}{#2}}}

\def\Half{{\fracd{1}{2}}}

\def\d{\partial}

\def\sign{\mathop{\rm sign}\nolimits}

\def\const{{\rm const}}
\def\eps{\varepsilon}
\def\ep{\epsilon}

\newcommand{\vars}[1]{\left\{\begin{array}{ll}#1\end{array}\right.}

\def\mN{_\mu^\nu}

\def\ep{\epsilon}

\def\wh{wormhole}
\def\whs{wormholes}

\def\sph{spherically symmetric}
\def\ssph{static, spherically symmetric}
\def\asflat{asymptotically flat}
 
\def\Schr{Schr\"o\-din\-ger}
\def\Scw{Schwarz\-schild}

\def\Veff{V_{\rm eff}}

\def\rf#1{\eqref{#1}}

\begin{document}
\onecol

\Title{The simplest wormhole in Rastall and k-essence theories}

\Author{Kirill A. Bronnikov\foom 1}
	{\small
	Center fo Gravitation and Fundamental Metrology, VNIIMS, 
		Ozyornaya ul. 46, Moscow 119361, Russia;\\
	 Institute of Gravitation and Cosmology, Peoples' Friendship University of Russia (RUDN University), 
		ul. Miklukho-Maklaya 6, Moscow 117198, Russia;\\
	National Research Nuclear University ``MEPhI'', 
		Kashirskoe sh. 31, Moscow 115409, Russia}
		
\Author{Vinícius A.G. Barcellos and Laura P. de Carvalho}
 	{\small N\'ucleo Cosmo-ufes\ \&\ Departamento de F\'{\i}­sica, CCE,
	Universidade Federal do Esp\'{\i}rito Santo,\\
	Vit\'oria, ES, CEP29075-910, Brazil}
 	   
\Author{J\'ulio C. Fabris\foom 2}
 	{\small N\'ucleo Cosmo-ufes\ \&\ Departamento de F\'{\i}sica, CCE,
	Universidade Federal do Esp\'{\i}­rito Santo,\\
	Vit\'oria, ES, CEP29075-910, Brazil\\
	National Research Nuclear University ``MEPhI'', 
		Kashirskoe sh. 31, Moscow 115409, Russia}
 
\Abstract
  {The geometry of the Ellis-Bronnikov wormhole is implemented in the Rastall and k-essence theories 
  of gravity with a self-interacting scalar field. The form of the scalar field potential is determined 
  in both cases. A stability analysis with respect to spherically symmetric time-dependent perturbations
  is carried out, and it shows that in k-essence theory the wormhole is unstable, like the original version 
  of this geometry supported by a massless phantom scalar field in general relativity. In Rastall's theory, 
  it turns out that a perturbative approach reveals the same inconsistency that was found 
  previously for black hole solutions: time-dependent perturbations of the static configuration prove 
  to be excluded by the equations of motion, and the wormhole is, in this sense, stable under 
  spherical perturbations. }

\email 1 {kb20@yandex.ru}
\email 2 {julio.fabris@cosmo-ufes.org}


\section{Introduction}

  Black holes and wormholes are remarkable predictions of the General Relativity theory (GR). The 
  detection of gravitational waves emitted by merging of compact objects \cite{ligo} and the recent 
  image of a supermassive object at the center of the galaxy M87 \cite{eht} have brought black holes 
  to the status of astrophysical objects whose existence in nature leaves little doubt. On the other
  hand, wormholes remain a hypothetical prediction of GR. In its simplest configuration, a wormhole is 
  composed of two asymptotically flat Minkowskian space-times connected by a kind of tunnel. The two flat 
  asymptotic regions are usually considered as different universes that are connected by a throat. One of 
  the problematic aspects of wormhole configurations is the necessity of having negative energy, at least 
  in the vicinity of the throat, in order that they could exist. Negative energy, which implies violation of the 
  standard energy conditions, brings two main problems: the configuration can be unstable; or generally, 
  the throat may not be traversable, in the sense that tidal forces may be huge, and possibly only pointlike 
  objects may cross it from one universe to the other, except for some special cases. For a 
  pedagogical description of wormhole properties, see Ref. \cite{thorne}.

  The Ellis-Bronnikov (EB) wormhole \cite{h_ell,kb73} is one of the simplest solutions of GR leading to 
  a structure of two flat asymptotics connected by a throat. As a matter content, the EB wormhole solution 
  uses a free massless scalar field with negative energy. Such field is normally denoted as a phantom scalar 
  field. The configuration is, as could be expected, unstable due to the repulsive nature of the scalar field, 
  see, e.g., \cite{zbf, ggs08} and references therein. Studies of static, spherically symmetric configurations 
  in the presence of scalar fields have a long history, see  \cite{fish,berg57} for the first seminal works on 
  these lines. In parallel, there has been much effort to obtain wormhole solutions which, besides being 
  traversable, would be stable and do not require exotic matter. 
  However, it is hard to fulfill these requirements in the context of GR and even in its extensions for a simple 
  reason: in order to cross the throat by coming from one region and arriving in the other, the geodesics must 
  first converge and later diverge, and this property requires repulsive properties of matter which should 
  thus violate at least some of the standard energy conditions. Still in the framework of GR there 
  are, on the one hand, an example \cite{kb13} of a stable wormhole supported by some kind of phantom 
  matter, and, on the other hand, examples of phantom-free rotating cylindrically symmetric \whs\ whose 
  stability properties are yet unknown \cite{kb-rot1, kb-rot2}. 

It is well known that a given metric may be a solution of the field equations of different theories of gravity
  or even in a single theory with different matter sources. An example is \cite{kb13} where the EB \wh\
  in GR is supported by a particular kind of phantom perfect fluid instead of a scalar field as in \cite{h_ell, kb73}.
  In the case of different theories, the matter content should naturally depend on the theory under consideration. 
  In this paper we explore the EB wormhole metric in two different theories. The first one is Rastall's theory
  of gravity \cite{prastall} that abandons one of the cornerstones of GR, the usual conservation law for matter 
  fields. The second one is the k-essence theory \cite{kess} which modifies the matter sector by introducing 
  non-canonical forms for the kinetic term of a scalar field. The k-essence proposal may be connected with 
  some fundamental theories inspired by quantum gravity. In both cases our goal is to verify if it is possible 
  to avoid the usual difficulties in wormhole construction and to obtain stable solutions.

  Previously, both Rastall and k-essence theories with a self-interacting scalar field have been studied in 
  attempts to obtain static, spherically symmetric black hole solutions \cite{rastall,k-ess}. The solutions 
  turned out to be quite exotic, mainly due to the asymptotic properties at infinity. A stability analysis has 
  shown that those k-essence solutions were unstable \cite{stab}. However, surprisingly, the perturbation
  analysis of the Rastall solutions was shown to be inconsistent, and the stability issue remained unclear 
  \cite{dk}. It has been speculated that this property of the Rastall solutions is connected with the absence 
  of a Lagrangian formulation of this theory. A curious aspect of these k-essence and Rastall solutions is 
  that they share some duality properties, in spite of quite different structures of the theories themselves 
  \cite{dual}.

  Here we show that the EB wormhole metric can be a solution of both Rastall and k-essence theories under 
  the condition that the potential describing the self-interaction of the scalar field is nonzero. We determine 
  the form of this potential in each case. In the k-essence theory we use a power-law expression of the 
  kinetic term, as in Ref. \cite{k-ess}. We perform a perturbation analysis of these solutions using a 
  gauge-invariant approach, and we find that the k-essence solution is unstable. Unlike that, in Rastall gravity 
  the inconsistency found previously for black hole solutions re-appears here, and no 
  time-dependent \sph\ perturbations can exist. Thus the EB metric in this framework may be said to be 
  stable under such perturbations, but the existence of nonperturbative time-dependent solutions cannot be
  excluded, to say nothing of possible instabilities under less symmetric perturbations.

  The paper is organized as follows. In Section 2 some general expression to be used in the calculations
  are settled out. In Section 3, the EB wormhole solution in GR is reproduced for comparison. In Section 4,
   the corresponding wormhole solution and the stability issue is presented for Rastall gravity. A similar 
   analysis is carried out in k-essence theory in Section 5. In Section 6 we present our conclusions.

\section{General relations}

  The goal of the present section is to give some general relations that will be used in the rest of the paper. 
  We assume spherical symmetry but not necessarily static. This allows us to easily consider a static 
  configuration which we will call the background and linear perturbations around it.

  Spherical symmetry can be described by a metric of the form
\bear           \label{m1}
		ds^2 = e^{2\gamma(t,x)}dt^2 - e^{2\alpha(t,x)}dx^2 - e^{2\beta(t,x)}d\Omega^2,
\ear
  where $d\Omega$ is the metric on a unit 2-sphere. If the configuration besides being
  spherically symmetric is also static, the metric coefficients $\alpha, \beta$ and $\gamma$ 
  depend only on the radial coordinate $x$. There is freedom to reparametrize the radial 
  coordinate, and its particular choice can be made by postulating a condition 
  connecting the coefficients $\alpha$, $\beta$ and $\gamma$.

  For the metric (\ref{m1}) the components of the Ricci tensor and expression for the 
  d'Alambertian operator acting on a scalar field are given by
\bear                             \label{Ric} 
	R^0_0 &=& e^{-2\gamma} \big[\ddot\alpha + 2\ddot\beta +
	          \dot\alpha{}^2 + 2 \dot\beta{}^2 - \dot\gamma(\dot\alpha + 2\dot\beta)]
		 - e^{- 2\alpha}\big[ \gamma''+ \gamma'(\gamma' - \alpha' + 2\beta')\big],
\nnv                 
	R^1_1 &=&   e^{ - 2\gamma}
        \big[\ddot\alpha + \dot\alpha(\dot\alpha - \dot\gamma + 2\dot\beta)\big] 
	          - e^{-2\alpha} \big[\gamma'' + 2\beta'' - \alpha'(\gamma'+ 2\beta') + \gamma'^2 
                                          + 2\beta'^2 \big] ,
\nnv          
	R^2_2 &=& e^{-2 \gamma}\big[\ddot\beta  
					+ \dot\beta(\dot\alpha - \dot\gamma + 2\dot\beta)\big] 
			- e^{-2\alpha}\big[\beta'' + \beta'(\gamma' - \alpha'+ 2\beta')\big] + e^{-2\beta},
\nnv         
	R_{01} &=&   2\dot\beta' + 2(\beta'- \gamma')\dot\beta - 2\beta'\dot\alpha ,
\yy					\label{Box}
	\Box\phi &=& e^{-2\gamma}\big[\ddot\phi + (\dot\alpha - \dot\gamma + 2\dot\beta)\dot\phi\big] 
		- e^{-2\alpha}\big[\phi'' + (\gamma' - \alpha' + 2\beta')\phi'\big].
\ear
  where dots denote $\d/\d t$ and primes $\d/\d x$.

  In the case of a static space-time, all time derivatives disappear. However, in the study 
  of small time-dependent perturbations around a given static solution at linear order, the 
  linear terms with time derivatives become relevant.

  In what follows we will discuss wormhole configurations in GR, Rastall's theory of gravity 
  in the presence of a scalar field, and k-essence theories. In all these cases, the gravitational 
  field equations can be written as the Einstein equations with appropriate stress-energy tensors
  $T\mN$,
\beq        \label{EE-G}
		R\mN - \frac{1}{2}\delta\mN R = - T\mN,
\eeq  
  or alternatively,
\bear        \label{EE-R}
		R\mN = - T\mN + \Half \delta\mN T_\rho^\rho,
\ear
  where we are using units in which (in usual notations) $c = 8\pi G =1$. These expressions are 
  also valid in Rastall gravity, under a suitable redefinition of the stress-energy tensor.

\section{Wormhole solution in GR with a free scalar field}

\subsection{The (anti-)Fisher solution and the simplest wormhole}

  Let us begin with recalling a derivation of the Ellis-Bronnikov wormhole solution in the context 
  of GR. The equations in the presence of a free massless scalar field $\phi$ are given by
\bearr        \label{EE1}
	R\mN - \frac{1}{2}\delta\mN R = 
		-\epsilon\bigg(\phi_{;\mu}\phi^{;\nu} - \frac{1}{2}\delta\mN \phi_{;\rho}\phi^{;\rho}\bigg),
\\ \lal           \label{e-phi1}  
		\Box\phi = 0,
\ear
   where the parameter $\epsilon$ indicates if the scalar field is of ordinary (canonical) 
   ($\epsilon = 1$) or phantom ($\epsilon = - 1$) type. The Einstein equations rewritten in the form 
   \rf{EE-R} read
\bear        \label{EE2}
		R\mN = -\epsilon\phi_{;\mu}\phi^{;\nu}.
\ear

  Let us consider the static metric \rf{m1} and a scalar field $\phi = \phi(x)$.
  The set of equations \rf{EE1} and \rf{e-phi1} is then most conveniently solved using 
  the harmonic coordinate condition $\alpha = 2\beta + \gamma$ \cite{kb73}
  (under which we will denote the radial coordinate by $u$). Indeed, under this condition, 
  the scalar field equation \rf{e-phi1} and two independent equations among \rf{EE2} 
  (specifically, $R^0_0 =0$ and $R^0_0 + R^2_2=0$) take the form 
\bearr
	  \phi'' =0, \qquad	\gamma'' =0, \qquad  \beta'' + \gamma'' = e^{2\beta+2\gamma}, 
\ear  
  (the prime stands here for $d/du$). All of them are immediately integrated giving 
\bearr                          \label{sol1}
		\phi = C u,\qquad  \gamma = - mu,  \qquad m, C = \const, 
\ear  
  (where two other integration constants are suppressed by choosing the scale of $t$ and 
  the zero point of $\phi$), and
\bearr                  \label{int1}
		(\beta' + \gamma')^2 = e^{2\beta+2\gamma} + k^2 \sign k, \qquad  k = \const, 
\ear  
   where one more integration constant is suppressed by choosing the zero point of $u$.
   The solution of \rf{int1} depends on the sign of $k$:
\bear
		e^{-\beta-\gamma} = k^{-1} \sinh (ku), && k > 0,
\nn		
		e^{-\beta-\gamma} = u, \cm\cm  && k=0,
\nn		
		e^{-\beta-\gamma} = k^{-1} \sin (ku), \ \ && k < 0,
\ear   
  which can be jointly written as 
\bearr                      \label{def-s}
		e^{-\beta-\gamma} = s(k,u) \equiv \vars{ k^{-1} \sinh (ku), & k > 0,\\
														 u, & k=0,\\
												      k^{-1} \sin (ku), & k < 0.}
\ear  
  Lastly, substituting \rf{sol1} and \rf{int1} into the ${1\choose 1}$ component of \eqs \rf{EE1} 
  (which is an integral of other components), we obtain a relation between the integration 
  constants:
\bear                         \label{int1a}
			k^2 \sign k = m^2 + \Half \epsilon C^2.
\ear  
  The metric takes the form 
\bear					\label{ds1}
		ds^2 = e^{-2mu} dt^2 - \frac{e^{2mu}}{s^2(k,u)}\bigg(\frac{du^2}{s^2(k,u)} + d\Omega^2\bigg),
\ear  
  The constants $m$ and $C$ have the meaning of the \Scw\ mass and the scalar charge, respectively.
  The coordinate $u$ is defined (without loss of generality) at $u >0$, and $u =0$ corresponds to 
  spatial infinity (since there $r(u) \equiv e^\beta \to \infty$), at which the metric is \asflat. 

  Equations \rf{sol1}, \rf{int1a} and \rf{ds1} give a joint representation of all \ssph\ solutions
  to \eqs \rf{EE1}, \rf{e-phi1}: Fisher's solution \cite{fish} of 1948 (repeatedly rediscovered
  afterwards)  corresponding to $\ep =1$ (hence $k > 0$) and all three branches of the 
  solution for $\ep = -1$ \cite{berg57} according to the signs of $k$ (sometimes called 
  anti-Fisher solutions). Detailed descriptions of the corresponding geometries can be found, 
  e.g., in \cite{kb73, bbook12, zbf, kb18}. Note that the instability of Fisher's solution 
  under small radial perturbations was shown in \cite{kb-hod}, and that of anti-Fisher
  solutions in \cite{ggs08, zbf}.
 
  Our interest here is with wormhole solutions, which form the branch $\ep = -1,\ k <0$: in this case,
  we have two flat spatial infinities at $u=0$ and $u = \pi/|k|$. The solution looks more transparent 
  after the radial coordinate transformation 
\beq
			x = b \cot (bu),  \qquad  b := |k|,  
\eeq  
  which brings the solution to the form 
\bearr 
 		     ds^2 = e^{-2m[\pi/2 - \arctan (x/b)]} dt^2 
 		     				- e^{2m[\pi/2 - \arctan (x/b)]} dx^2 - (x^2 + b^2) d\Omega^2,
\yyy
			\phi = C \big [\pi/2 - \arctan (x/b)\big],		     				
\ear 
  where $x$ is the so-called quasiglobal coordinate corresponding to the ``gauge'' $\alpha + \gamma =0$
  in terms of the metric \rf{m1}. The simplest configuration is obtained in the case of zero mass, $m = 0$:
\bearr             \label{m-ell}
		ds^2 = dt^2 - dx^2 - (x^2 + b^2) d\Omega^2, 
\yyy           \label{phi1}
		\phi =\pm b \sqrt{2} \big [\pi/2 - \arctan (x/b)\big].	
\ear
  It is this solution that is called the Ellis \wh\ \cite{h_ell}, or the Ellis-Bronnikov (EB) \wh, since this and 
  more general scalar-vacuum and scalar-electrovacuum configurations were obtained and discussed
  in \cite{kb73}. In terms of the metric \rf{m1}, we have in \rf{m-ell}
\beq                   \label{back}
		\alpha \equiv \gamma \equiv 0, \qquad  \beta \equiv \log r(x) = \Half \log (x^2 + b^2).
\eeq  

\subsection{Ellis \wh\ instability in GR}

  Consider now linear time-dependent \sph\ perturbations of the EB \wh, described by additions 
  $\delta\alpha$, $\delta\beta$, $\delta\gamma$ and $\delta\phi$ of the corresponding static
  (background) quantities, characterized by some smallness parameter $\eps$. 
  Following \cite{kb-hod, zbf, kb18}, we choose the perturbation gauge $\delta\beta =0$. Then 
  the perturbation equations following from \rf{e-phi1} and \rf{EE1} in the order $O(\eps)$
  can be written as
\bear          \label{e-phi1a}
	 e^{2(\alpha - \gamma)}\delta\ddot\phi - \delta\phi'' - [2\beta'+ \gamma' - \alpha' ]\delta\phi' 
	 		- [\delta\gamma'- \delta\alpha']\phi' \eql 0.
\yy				\label{00-1a}
	e^{2(\alpha - \gamma)}\delta\ddot\alpha - \delta\gamma'' 
		- \delta\gamma'(2\gamma' - \alpha' + 2\beta') + \gamma' \delta\alpha' \eql 0,
\yy				\label{11-1a}
	e^{2(\alpha - \gamma)}\delta\ddot\alpha  - \delta\gamma'' + \delta\alpha'(\gamma'+ 2\beta') 
		+ (\alpha' - 2\gamma')\delta\gamma'  \eql  2\epsilon\phi'\delta\phi' ,
\yy				\label{22-1a}
	\beta'(\delta\gamma' - \delta\alpha')  - 2e^{2(\alpha - \beta)} \delta\alpha  \eql 0,	
\yy				\label{01-1a}		
	- \beta'\delta\dot\alpha \eql - \frac{\epsilon}{2}\phi'\delta\dot\phi.
\ear
  These equations are written with an arbitrary radial coordinate $x$ in the background static metric 
  but with a particular choice ($\delta\beta=0$) of the perturbation gauge fixing the reference 
  frame in perturbed space-time. We see that \eq\rf{01-1a} can be integrated in $t$ giving
\bear          \label{da1}
		\delta\alpha = - \frac{\eta}{2}\delta\phi + \xi(x), \qquad \eta = \frac{\phi'}{\beta'},
\ear
  with an arbitrary function $\xi(x)$; we will put $\xi(x) \equiv 0$ since only time-dependent
  perturbations are of interest. 
   
  For the Ellis wormhole solution \rf{m-ell}, \rf{phi1}, such that $\gamma = \alpha = 0$ and 
  $\ep = -1$, the remaining equations read
\bear  		\label{e-phi1b}
		\delta\ddot\phi - \delta\phi'' - 2\beta'\delta\phi' - \phi' (\delta\gamma'- \delta\alpha') \eql 0.
\yy		\label{grpe-1}
		\delta\ddot\alpha - \delta\gamma'' - 2 \beta' \delta\gamma' \eql 0,
\yy			\label{grpe-2}
		\delta\ddot\alpha - \delta\gamma'' + 2\beta'\delta\alpha'  = - 2\phi'\delta\phi' ,
\yy			\label{grpe-4}
		\beta'(\delta\gamma' - \delta\alpha')  - 2e^{- 2\beta} \delta\alpha  = 0,
\ear
  Subtracting equations (\ref{grpe-1}) and (\ref{grpe-2}) and using \rf{da1}, we obtain
\bear  
      	\delta\gamma' = \frac{1}{2}(\eta'\delta\phi - \eta\delta\phi').
\ear
  Knowing $\delta\alpha$ and $\delta\gamma'$, or equivalently using directly \rf{grpe-4},
  we can eliminate the metric perturbations from the scalar field equation, which results in
  the following master equation:
\bear           \label{mast0}
		\delta\ddot\phi - \delta\phi'' - 2\beta'\delta\phi' - \eta'\phi' \delta\phi = 0.
\ear
  Assuming the time dependence of $\delta\phi$ as a single spectral mode, 
  $\delta\phi \propto e^{i\omega t}$, 
\bear           \label{mast0a}
		\delta\phi''+ 2\beta'\delta\phi' +(\omega^2 + \eta'\phi')\delta\phi = 0.
\ear
   Eliminating $\delta\phi' $ by the substitution $\delta\phi = e^{-\beta}y (x)$, we arrive at
   the \Schr\--like equation
\bear  			\label{mast1}
		y''+ \biggr\{\omega^2 + \eta'\phi' - \beta'' - \beta'^2\biggl\}y = 0,
\ear
  which coincides with the master equation found in \cite{zbf} in the special case where 
  $\alpha = \gamma = 0$ and no scalar field potential is present. Using our expressions
  for $\phi$ and $\beta$ in the Ellis \wh\ solution, we find
\bear  			\label{mast1a}
		y'' + \biggr\{\omega^2 - \biggr[\frac{b^2(3x^2 + 2b^2)}{x^2(x^2 + b^2)^2}\biggl]\biggl\}y = 0.
\ear
  The stability analysis requires imposing boundary condition. In our case, for 
  $x \to \pm \infty$ it is reasonable to require $\delta\phi \to 0$, or $y = o (|x|)$. 
  We can note that, asymptotically, \eq \rf{mast1a}) has solutions in terms of Bessel functions,
\bear  
		y (x) = A_{\pm}\sqrt{|x|}J_{\pm\nu}(\omega |x|), \qquad \nu = \sqrt{3b^2 + 1/4},
		\qquad A_\pm = \const.
\ear		
  If $\omega = i\bar\omega$ (an imaginary frequency describing an instability), the solutions 
  become
\bear  
		y(x) = \sqrt{|x|}\Big[A_1 K_\nu(\bar\omega|x|) + A_2 I_\nu(\bar\omega |x|)\Big],
		\qquad A_{1,2} = \const.
\ear
  where $K_\nu$ and $I_\nu$ are modified Bessel functions. The function $K_\nu$ tends to 
  zero at large $|x|$, therefore, correct boundary conditions with imaginary $\omega$ are
  compatible with an instability. On the other hand, the positive nature of the effective potential
  $\Veff(x)$ in \eq\rf{mast1a} (the expression in brackets) seems to exclude ``energy levels'' 
  $\omega^2 < 0$. However, this argument cannot be directly applied because of a pole  
  of this effective potential near the \wh\ throat $x = 0$, $\Veff \approx 2/x^2$.  This potential 
  can be regularized by the appropriate Darboux transformation as described in 
  \cite{ggs08, zbf}. The regularized potential turns out to contain a sufficiently deep well leading
  to the existence of an unstable perturbation mode, related to an evolving throat radius. 
  The same result was previously obtained by a numerical study \cite{shinkai} which proved
  that an Ellis \wh\ can either collapse to a black hole or inflate, depending on the sign 
  of the initial perturbation.
  
The gauge condition we are using, $\delta\beta =0$, seems to prevent considering 
  perturbations connected with a changing throat radius. But a more thorough investigation 
  shows \cite{ggs08, zbf} that the unknown $\delta\phi$ in the master equation \eq\rf{mast0} 
  is actually a gauge-invariant quantity. Indeed, a perturbation gauge may be described as 
  a small coordinate shift $x^\mu \to x^\mu + \xi^\mu$ with $\xi^\mu = O(\eps)$, or, in the 
  ($x,t$) subspace,
\[
		t = \bar t + \xi^0 (x,t), \qquad   x = \bar x + \xi^1 (x,t).
\]  
Then it can be directly verified that quantities like $\beta'\delta\phi - \phi' \delta\beta$
  do not change under such coordinate shifts and are thus gauge-invariant, as well as their 
  products with any background quantities, for example, $1/\beta'$. It follows that 
  $\delta\phi$ in our consideration is the specific form of the gauge-invariant quantity 
  $\psi = \delta\phi - \phi' \delta\beta'/\beta'$ in the gauge $\delta\beta =0$. Other functions
  involved in \rf{mast0} are combinations of the background quantities, therefore we can safely 
  replace there $\delta\phi$ with $\psi$ and conclude that the whole master equation is
  gauge-invariant. It can thus be used for considering any perturbations, including those
  with an evolving throat radius.
  
The gauge invariance issue is presented in more detail in \cite{ggs08, zbf, kb18}, and its
  analogue for perturbations in cosmology is discussed in \cite{brand}. In our further 
  consideration we obtain gauge-invariant master equations for spherical perturbations in 
  a similar way. 

\section{Wormholes in Rastall gravity}

  In Rastall's theory, if the source of gravity is a scalar field $\phi$ with a self-interaction 
  potential $V(\phi)$, the field equations can be written as \cite{rastall, dk}
\bearr         \label{EE-G2}
		R\mN - \frac{1}{2}\delta\mN R = - \epsilon\biggr\{\phi_{;\mu}\phi^{;\nu} 
		+ \frac{2 - a}{2}\delta\mN \phi^{;\rho}\phi_{;\rho}\biggl\} - (3 - 2a)\delta\mN V(\phi),
\yyy             \label{e-phi2}
	\Box\phi + (a - 1) \frac{\phi^{;\rho}\phi^{;\sigma}\phi_{;\rho;\sigma}}
			{\phi_{;\alpha}\phi^{;\alpha}} = - \epsilon(3 - 2a)V_\phi,
\ear
   where $a$ is a constant parameter of the theory, and at its special value $a=1$ 
   we return to GR. Thus the effective stress-energy tensor of the scalar field reads
\beq   
		T\mN = \epsilon\biggr\{\phi_{;\mu}\phi^{;\nu} 
		  - \frac{2-a}{2}\delta\mN \phi^{;\rho}\phi_{;\rho}\biggl\} +\delta\mN W(\phi),   
\eeq   
   where $W(\phi) = (3-2a) V(\phi)$. The modified Einstein equations can be rewritten as
\bear                   \label{EE-R2}
		R_{\mu\nu} = -\epsilon\biggr\{\phi_{;\mu}\phi_{;\nu} 
			+ \frac{1 - a}{2}g_{\mu\nu}\phi^{;\rho}\phi_{;\rho}\biggl\} + g_{\mu\nu}W(\phi).
\ear

  For the static metric \rf{m1} and $\phi = \phi(x)$, the Rastall equations reduce to
\bear          \label{e-phi2a}
	a\phi'' + [\gamma'- a\alpha' + 2\beta']\phi' \eql \epsilon  e^{2\alpha}W_\phi,
\yy                      \label{00-2a}
	\gamma''+ \gamma'(\gamma' - \alpha' + 2\beta') \eql 
			- \frac{\epsilon}{2}(1 - a)\phi'^2 - e^{2\alpha}W ,
\yy                      \label{11-2a}			
	\gamma'' + 2\beta'' - \alpha'(\gamma'+ 2\beta') + \gamma'^2 + 2\beta'^2 
			\eql - \frac{\epsilon}{2}(3 - a)\phi'^2 - e^{2\alpha}W,
\yy	                      \label{22-2a}
	\beta'' + \beta'(\gamma' - \alpha'+ 2\beta') - e^{2(\alpha - \beta)} 
		\eql - \frac{\epsilon}{2}(1 - a)\phi'^2 - e^{2\alpha}W,
\ear
  where $W_\phi = dW/d\phi$.
  
  These equations become identical to the GR equations with a massless scalar field if
  we put
\bear  		\label{pot1}
		\frac{\epsilon}{2}(1 - a)\phi'^2  = - e^{2\alpha}W.
\ear
  where we should take into account that $W_\phi = W'/\phi'$.
  Then all solutions for $\alpha$, $\beta$, $\gamma$ and $\phi'$ are the same as in GR. 
  But, a new element in Rastall gravity is that one needs a nonzero potential in order to create 
  these solutions. For any given special solution, the potential can be determined from 
  \eq\rf{pot1} or from any of the equations \rf{00-2a}--\rf{22-2a}.
  
  In particular, for the Ellis \wh\ solution \rf{m-ell}, \rf{phi1} the potential is found to be  
\bear  				\label{W}
	           W(\phi) \equiv (3-2a) V(\phi) = \frac{b^2 (1-a)}{(x^2 +b^2)^2}
	           		= \frac{1-a}{b^2} \ \cos^4 (\phi/\sqrt 2). 
\ear
  Thus we have the simplest Ellis \wh\ solution in Rastall gravity for any value of 
  the Rastall parameter $a$.

\subsection{Wormhole stability in Rastall gravity}
 
  To obtain the linear perturbation equations, we are consider \eqs \rf{e-phi2} and \rf{EE-R2}
  using the expressions \rf{Ric} for the Ricci tensor components,
  the gauge $\delta\beta = 0$ and the potential \rf{W} as a function of $\phi$.
  The equations read
\bearr      		\label{e-phi2b}
		e^{2(\alpha - \gamma)}\delta\ddot\phi - a\delta\phi'' - (\gamma'- a\alpha' + 2\beta')\delta\phi' 
		- \phi'(\delta\gamma'- a\delta\alpha') 
		= -\ep e^{2\alpha}(2 W_\phi \delta\alpha + W_{\phi\phi}\delta\phi),
\yyy		
		e^{2(\alpha - \gamma)}\delta\ddot\alpha - \delta\gamma''
		- (2\gamma' - \alpha' + 2\beta')\delta\gamma' + \gamma' \delta\alpha'
		= \ep(1 - a)\phi'\delta\phi' +  e^{2\alpha}(2 W \delta\alpha + W_\phi\delta\phi),
\yyy
		e^{2(\alpha - \gamma)}\delta\ddot\alpha
		- \delta\gamma'' + (\gamma'+ 2\beta')\delta\alpha' + (\alpha' - 2\gamma')\delta\gamma' 
		 = \ep(3 - a)\phi'\delta\phi' +  e^{2\alpha}(2 W \delta\alpha + W_\phi\delta\phi),
\yyy
		\beta'(\delta\gamma' - \delta\alpha')  - 2e^{2(\alpha - \beta)} \delta\alpha 
		 = - \ep(1 - a)\phi'\delta\phi' -  e^{2\alpha}(2 W \delta\alpha + W_\phi\delta\phi),
\yyy
		- \beta'\delta\dot\alpha =- \frac{\epsilon}{2}\phi'\delta\dot\phi. 
\ear
   For  our simplest case $\gamma = \alpha = 0$, $\ep = -1$, the equations read 
\bearr    		\label{e-phi2c}
		  \delta\ddot\phi  - a\delta\phi'' - 2\beta'\delta\phi' -\phi'(\delta\gamma'- a\delta\alpha')
			 = -\ep (2 W_\phi \delta\alpha + W_{\phi\phi}\delta\phi),
\yyy                  \label{pr1}
      \delta\ddot\alpha - \delta\gamma'' + 2\beta'\delta\gamma'=
			\epsilon(1 - a)\phi'\delta\phi' +  (2 W \delta\alpha + W_\phi\delta\phi),
\yyy                       \label{pr2}
		\delta\ddot\alpha - \delta\gamma'' - 2\beta'\delta\alpha' 
			= \epsilon(3 - a)\phi'\delta\phi' +  (2 W \delta\alpha + W_\phi\delta\phi),
\yyy                      \label{pr4}
		\beta'(\delta\gamma' - \delta\alpha')  - 2e^{- 2\beta} \delta\alpha  =
			- \epsilon(1 - a)\phi'\delta\phi' -  (2 W \delta\alpha + W_\phi\delta\phi),
\yyy			\label{pr3}
		 \beta'\delta\dot\alpha = - \frac{1}{2}\phi'\delta\dot\phi.
\ear

  In  \cite{dk} it has been shown that the stability problem for the Rastall theory in 
  static, spherically symmetric configurations is inconsistent unless all perturbations are zero. 
  It turns out that here we come across the same problem, as could be expected 
  in view of those results. Indeed, from \eq (\ref{pr3}) we obtain, as previously in GR,
\bear                 \label{pr-a}
		\delta\alpha = - \Half \eta\delta\phi, \qquad \eta = \frac{\phi'}{\beta'}.
\ear
  From the difference of (\ref{pr1}) and (\ref{pr2}) we obtain
\bear  
		\delta\gamma'= \frac{\epsilon}{2}(\eta\delta\phi' - \eta'\delta\phi).
\ear
  On the other hand, from \eqs (\ref{pr4}) and (\ref{pr-a}) it follows
\bear                        \label{pr-b}
	\delta\gamma' = - \Half \Big( \eta\delta\phi' - \eta'\delta\phi\Big) 
	+(1 - a)\eta\bigg[\delta\phi' + \bigg(\frac{\eta}{4}\phi' + \frac{\phi''}{\beta'}\bigg)\delta\phi\bigg].
\ear
  In this expression we have separated the terms contained in (\ref{pr-a}) from the others.
 
  The expressions (\ref{pr-a}) and (\ref{pr-b}) coincide only if $a = 1$, that is, when the Rastall 
  theory reduces to GR, or if the quantity in brackets in (\ref{pr-b}) vanishes. In the second case, 
  we can find explicitly the behavior of $\delta\phi$:
\bear                   \label{pr-c}
               \delta\phi = \phi_1(t) \exp \bigg(\! - \frac{3}{2}x + 2b\arctan\frac{x}{b}\bigg),
               \qquad   \phi_1(t) = \mbox{arbitrary function}.
\ear
  It is easy to see that, according to \eqs(\ref{e-phi2c}) and (\ref{pr-a}), in the solution (\ref{pr-c}) 
  the only possibility is $\phi_1(t) =0$. Hence, there is no perturbation at linear level, the same result
  as in \cite{dk}. Quite similarly to \cite{dk}, it implies the absence of perturbations in all 
  orders of smallness.
  
\section{Wormholes in k-essence theories}
\subsection{Static wormholes}

  Let us consider the theory defined by the Lagrangian density
\bear  
		{\cal L} = \sqrt{-g}\Big[R + f(X) - 2V(\phi)\Big],
\ear
  with the definitions
\bear  
		X = \eta \phi_{;\rho}\phi^{;\rho}, \qquad \eta = \pm 1.
\ear
  The scalar field equation has the form 
\bear               \label{e-phi3}
		\eta f_X\Box\phi + 2f_{XX}\phi^{,\rho}\phi^{,\sigma}\phi_{;\rho;\sigma} = V_\phi,
\ear		
  where the subscripts $X$ and $\phi$ denote derivatives with respect to the corresponding 
  variables. The Einstein equations have the form \rf{EE-G} with the stress-energy tensor
\beq
		T\mN = f_X \eta \phi_{,\mu}\phi^{,\nu} - \Half \delta\mN f + \delta\mN V.
\eeq  
  In the form \rf{EE-R} they can be written as 
\bear  	
		R\mN  = \eta f_X\phi_{,\mu}\phi^{,\nu} - \Half \delta\mN (- f + Xf_X + 2V).
\ear

  Let us now consider static, spherically symmetric space-times with the metric \rf{m1} 
  and $\phi = \phi(x)$ and choose
\bear  
		f(X) = \ep f_0 X^n,      \qquad   n >0, \qquad f_0>0, \qquad \ep = \pm 1.
\ear
  To avoid the possibility of complex values of $f(X)$, we must then fix $\eta = - 1$, 
  so that
\bear  
			X = e^{-2\alpha}\phi'^2.
\ear
  The resulting equations of motion are
\bear    		\label{e-phi3a}
	nf_0 e^{-2n\alpha}\phi'^{2n-2}\Big\{
		(2n - 1)\phi'' + [2\beta'+\gamma'- (2n - 1)\alpha' ]\phi'\Big\} \eql - \ep V_\phi,
\yy		             \label{00-3}
	\gamma''+ \gamma'(2\beta' + \gamma' - \alpha') =
				- \frac{\ep f_0}{2}(n - 1)e^{2(1-n)\alpha}{\phi'}^{2n} - e^{2\alpha}V,
\yy		             \label{11-3}				
	\gamma'' + 2\beta'' - \alpha'(\gamma'+ 2\beta') + \gamma'^2 + 2\beta'^2 
		= \frac{f_0}{2}(n + 1)e^{2(1-n)\alpha}\phi'^{2n} - e^{2\alpha}V,
\yy				   \label{22-3}
	\beta'' + \beta'(2\beta' + \gamma' - \alpha') - e^{2(\alpha - \beta)} 
			= - \frac{\ep f_0}{2}(n - 1)e^{2(1-n)\alpha}\phi'^{2n} - e^{2\alpha}V.
\ear
  
    If we assume that the Ellis wormhole is a solution to \eqs (\ref{e-phi3a})-(\ref{22-3}), 
    we substitute there the expressions \rf{back} and find that the sum and difference 
    of \rf{00-3} and \rf{11-3} leads to the relations
\bear                   	\label{kr1}
		V = \frac{\ep f_0}{2}(n-1) e^{2n\gamma}{\phi'}^{2n}, \qquad
		nf_0 \ep \phi{}^{2n} = \frac {2 r''}{r} = \frac{2 b^2}{(x^2 + b^2)^2}. 
\ear
   It follows that $\ep = -1$, which is natural for a \wh\ solution that must violate the 
   Null Energy Condition, so that $T^0_0 - T^1_1 < 0$. As a result, we obtain the following 
   explicit expressions for $\phi'$ and the potential $V$:
\bear         \label{phi'3}
		\phi' \eql C (x^2 + b^2)^{-1/n}, \qquad  C = \bigg(\frac{2b^2}{nf_0}\bigg)^{1/(2n)},
\yy              \label{V3}  
		V \eql \frac{f_0}{2} (1 - n) {\phi'}^{2n} = \frac{f_0}2 (1-n) \frac{C^{2n}}{(x^2+b^2)^2}. 
\ear  
   Substituting the expression for $\phi'$ to \rf{e-phi3a} to find $V_\phi$, one can verify
   that the latter coincides with $V_\phi = V'/\phi'$ obtained directly from \rf{V3}, thus
   confirming the correctness of the solution.

  One can integrate $\phi'$ given by \rf{phi'3} to obtain
\bear  
	\phi = Cx b^{-2/n}\,{ _2F_1}\Big(\frac12,\ \frac 1n,\ \frac 32;\ - \frac{x^2}{b^2}\Big) + \phi_0,
		\qquad \phi_0 = \const.
\ear
  It is not simple to obtain a closed expression for $V(\phi)$ (to do that, we must invert the  
  hypergeometric function). However, $V(\phi)$ is well defined since $\phi' > 0$ at all $x$.
  Also, at some special values of $n$ the hypergeometric function can reduce to simpler 
  expressions.

  Thus the Ellis wormhole solution is consistent with k-essence theory with a potential.

\subsection{Instability of the k-essence solution}

  The perturbation equations in the gauge $\delta\beta = 0$, under the condition
  $\alpha = \gamma =0$ (but their perturbations are nonzero) can be written as
\bearr     \label{e-df3}
		 \delta\ddot\phi  - (2n - 1)\delta\phi'' - 2\beta'\delta\phi' 
		 - \biggr\{\delta\gamma' - ( 2n - 1)\delta\alpha' \biggl\}\phi' 
\nnn \inch		
		 = \frac{1}{nf_0}{\phi'}^{(1 - 2n)}\biggr\{V_{\phi\phi}\phi'\delta\phi 
		 + 	V_\phi\biggr[2n\phi'\delta\alpha + 2(1 - n)\delta\phi'\biggl]\biggl\},
\yyy		 \label{pke1}
	\delta\ddot\alpha - \delta\gamma'' - 2\beta'\delta\gamma'
	 = f_0(n - 1){\phi'}^{2n - 1}\Big[(1- n)\phi'\delta\alpha + n\delta\phi'\Big]
			+ \Big(2 V \delta\alpha + V_\phi\delta\phi\Big),
\yyy			\label{pke2}
	 \delta\ddot\alpha	 - \delta\gamma'' - 2\beta' \delta\alpha'
	 = - f_0(n + 1){\phi'}^{2n - 1} \Big[ (1- n)\phi'\delta\alpha + n\delta\phi'\Big] 
				+ \Big(2 V \delta\alpha + V_\phi\delta\phi\Big),
\yyy                   \label{pke4}
	 \beta'(\delta\gamma' - \delta\alpha')  - 2e^{- 2\beta} \delta\alpha  = 
		f_0(1- n){\phi'}^{2n - 1}\Big[(1- n)\phi'\delta\alpha + n\delta\phi'\Big] 
		- \Big(2 V \delta\alpha + V_\phi\delta\phi\Big),
\yyy              \label{pke3}
		- \beta'\delta\dot\alpha = \frac{n}{2}f_0{\phi'}^{2n - 1}\delta\dot\phi.
\ear

   From \eq (\ref{pke3}) one obtains
\bear                   \label{pke6}
	\delta\alpha = - \frac{n}{2}\bar\eta\delta\phi, \qquad \bar\eta = f_0\frac{\phi'^{2n - 1}}{\beta'}.
\ear
  Using this result, and combining \eqs (\ref{pke1}) and (\ref{pke2}), we obtain
\bear                      \label{pke7}
	\delta\gamma' = \frac{n}{2}(1 - 2n)\bar\eta\delta\phi' 
	+ \frac{n}{2}\biggr[\bar\eta' + (1 - n)n\bar\eta^2\phi'\biggl]\delta\phi.
\ear
  This expression is consistent with (\ref{pr-b}) if $n = a = 1$ and $\ep = - 1$.

  Using \eq (\ref{pke4}), the relation (\ref{pke6}) and the background equations, we find again 
  \eq (\ref{pke7}). Hence, unlike the Rastall case, the k-essence perturbation analysis is 
  consistent, quite similarly to the results of \cite{stab,dk}.
  
  In addition, we can obtain an expression for $\delta\alpha'$ by combining \rf{pke4}
  with the difference of \rf{pke1} and \rf{pke2} as
\beq  
		\delta\alpha' = \frac 1{2\beta'} \Big[ -nf_0 \phi'{}^{2n-1}
				 - 2e^{-2\beta} \delta\alpha -2(1-n) f_0 \beta' \phi'{}^{2n-1} \delta\phi \Big].  
\eeq
  This expression coincides with the one obtained by directly differentiating \rf{pke6},
  which verifies the correctness of the model and the calculations.  

  Now, to obtain the master equation for $\delta\phi$, we use the previous results and 
  insert into them (\ref{e-df3}), along with the relations due to the background equations,
\bear
		 V_\phi \eql  - n\bar\eta\beta'\biggr[(2n - 1)\frac{\phi''}{\phi'} + 2\beta'\biggl]
		    = \frac {4b^2}{C n}\,(n-1) x (b^2 + x^2)^{-3+1/n},
\yy
	 V_{\phi\phi} \eql - n\biggr\{4\eta\beta''\frac{\beta'}{\phi'} + \bar\eta'\frac{\beta'}{\phi'}
		\biggr[(2n - 1)\frac{\phi''}{\phi'} + 2\beta'\biggl]  
		+ (2n - 1)\bar\eta\biggr[\frac{\beta''\phi''}{\phi'^2} 
		+ \beta'\biggr(\frac{\phi'''}{\phi'^2} - \frac{\phi''^2}{\phi'^3}\biggl)\biggl]\biggl\}
\nnv
			\eql   \frac  {4 b^2 (n-1)}{C^2 n^2}\, (b^2 + x^2)^{-4 + 2/n} (n b^2 + (2 - 5 n) x^2).
\ear
  The final form of the master equation is
\bearr                    \label{keme}
        - \delta\ddot\phi + (2n - 1)\delta\phi'' + \bigg\{2\beta' + 2\frac{(1 - n)}{n}\frac{V_\phi}{\bar\eta\beta'}\bigg\}\delta\phi' 
\nnn   \inch\inch
	+ \biggr\{n^2\bar\eta'\phi' +\frac{n(1 - n)}{2}\bar\eta^2\phi'^2 - n\frac{\phi'}{\beta'}V_\phi 
	     + \frac{\phi'V_{\phi\phi}}{n\bar\eta\beta'}\biggl\}\delta\phi = 0,
\ear
   or explicitly,
\bearr                    \label{master}
		 \delta\ddot\phi - (2n - 1)\delta\phi'' - 2\beta'\bigg[ 1 - \frac{2(n-1)^2}{n}\bigg] \delta\phi' 
		 			+ U(x)  \delta\phi =0,
\nnnv	\cm 			
			U(x) = \frac{2b^4 (2n-1)}{x^2 (x^2+b^2)^2} 
			          + \frac 2{n^2(x^2+b^2)^2} \Big[ n b^2 (n^2 + n -1) -x^2 (5n^2 - 7n +2)  \Big].
\ear   
   
  In general, the analysis of \eq \rf{master} is quite complicated. Let us begin with
  a simple example which shows a particular case where it is possible to explicitly prove 
  the instability. Particularly, let us fix $n = 1/2$. This case has been investigated in 
  \cite{k-ess,dk} in search for black hole solutions and a study of their stability. In fact, 
  there are some exotic types of black hole, but they are unstable. Now we are considering
  the same problem for the Ellis \wh.

  With $n = 1/2$, \eq \rf{master} greatly simplifies and reads
\beq
			\delta\ddot\phi - \frac{2x^2 + b^2}{(x^2+b^2)^2}\ \delta \phi = 0,			
\eeq  
  which is easily integrated giving
\beq                      \label{k-half}
				\delta\phi = K_1(x) e^{H(x)t} + K_2(x) e^{-H(x)t}, \cm 
							H(x) = \frac{\sqrt{2x^2 + b^2}}{x^2+b^2},	
\eeq  
  where $K_1(x)$ and $K_2(x)$ are arbitrary functions. This evidently demonstrates the instability 
  of the background configuration since the expression \rf{k-half} exponentially grows with time 
  if $K_1 \ne 0$.
  
  If $n=1$, we return to the situation in GR.

  If $ n < 1/2$, \eq\rf{master} loses its hyperbolic nature, and the system is hydrodynamically 
  unstable for the same reason as described in \cite{stab} and other papers. 
  
  Of more interest is the situation where $n > 1/2$, in which \eq\rf{master} has a wave nature.  
  It is then reasonable to get rid of the term containing $\delta\phi'$ by putting
\beq
			\delta\phi = e^{-p\beta} y(x,t), \cm   p = \frac{2n^2 - 5n +2}{n(1- 2n)}, 
\eeq  
  after which the equation acquires the form
\beq
			\ddot y - (2n-1) y'' + [U(x) + (2n-1) (\beta'' + p \beta'{}^2) ] y =0,
\eeq  
  or, assuming a single spectral mode, $y \propto e^{i\omega t}$, so that $\ddot y = - \omega^2 y$,
\beq               \label{Schr}
			y'' + \bigg[\frac{\omega^2}{2n-1} - V_{\rm eff}(x)\bigg] y =0,
			\cm
			V_{\rm eff}(x) = \frac{U(x)}{2n-1} + p\beta'' + p^2 \beta'{}^2.
\eeq  

  It is the Schr\"odinger-like equation usually appearing in stability studies, for which the corresponding 
  boundary-value problem should be solved in order to obtain stability conclusions. 
  For perturbations of wormholes with phantom scalar fields, the effective potentials $V_{\rm eff}(x)$ 
  contain a singularity on the throat which can be regularized with proper Darboux transformations
  \cite{zbf, ggs08, kb18} under the condition that $V_{\rm eff}(x) = 2/x^2 + O(1)$ near the throat 
  (where $x$ is the ``tortoise'' coordinate in the wormhole space-time, and $x=0$ is the throat).
  This condition is generally satisfied for wormholes supported by phantom scalar fields with arbitrary 
  potentials \cite{kb18}. Surprisingly, this condition also holds for the effective potential
  $V_{\rm eff}(x)$ in our equation \rf{Schr} for wormholes in k-essence theory, 
  so that the stability problem can be solved along the lines of  \cite{zbf, ggs08, kb18}.
  This requires a separate study, to be performed in the near future.

\section{Conclusion}

 The Ellis-Bronnikov solution represents the simplest analytical wormhole solution that can be obtained in 
 GR. It consists of two asymptotically flat regions connected by a throat. This wormhole requires a massless, 
 minimally coupled phantom scalar field: this means that all space, not only the throat, is filled with a field 
 having negative energy density. In spite of being a very elegant and simple solution, the EB wormhole 
 has a major drawback: it is unstable under linear perturbations. To look for a simple wormhole solution 
 like the EB one that may not require phantom fields and/or that are stable is a challenge, even if some 
 extensions of GR are employed. 
 
 It is well known that the same metric can be a solution of different gravitational theories. We have exploited 
 this fact in order to investigate the conditions at which the EB wormhole metric can be a solution in the 
 context of two extended gravity theories, Rastall gravity and k-essence. Rastall gravity is a more radical 
 departure from GR since it modifies the usual expression for the conservation law. In some sense, Rastall 
 gravity may be recast in the structure similar to GR where the expression for the energy-momentum tensor 
 must be modified. Unlike that, k-essence is essentially a modification of the matter sector, keeping a 
 Lagrangian formulation, by generalizing the usual kinetic expression. Both theories have applications, 
 for example, in cosmology \cite{daouda,kess,pert-k-ess} and black hole physics \cite{rastall,k-ess}.

 We have shown that the EB metric can be a static, spherically symmetric solution in both Rastall and 
 k-essence theories. To achieve that, a potential must be added in both cases, implying that, as opposed to GR, 
 a self-interacting scalar field is required. The next step was to investigate the stability of these solutions in 
 Rastall and k-essence cases. In Rastall gravity we face the same feature that was already found for 
 black hole solutions: the usual perturbative approach leads to inconsistencies forcing to set all fluctuations 
 near the background solution equal to zero. Perhaps this curious property is connected with the absence of 
 a Lagrangian formulation. In k-essence theory, we have shown that the wormhole is unstable with respect 
 to linear perturbation at least for the parameter $n$ in the range $n \leq 1/2$. 

 Two remarks must be added concerning these results of the perturbation analysis. First, only the simplest 
 version of a wormhole metric has been investigated. This restriction is motivated by technical reasons since 
 more complex configurations lead to very cumbersome expressions for the perturbations, even if a master 
 equation can be obtained. Very probably, numerical investigation can be necessary, which may imply new 
 technical challenges. However, this remark mainly concerns the k-essence case. In Rastall gravity the 
 situation may be more involved since we can expect that the inconsistency found here
 in the perturbative approach must remain, then a nonperturbative approach must be implemented. 
 We hope to address these problems in future studies.

\bigskip

\noindent
{\bf Acknowledgements:} 
  V.B., L.C. and J.F thank FAPES (Brazil) and CNPq (Brazil) for partial financial support. 
  K.B. was supported by the RUDN University Strategic Academic Leadership Program. The research of K.B. 
  was also funded by the Ministry of Science and Higher Education of the Russian Federation, Project
  ``Fundamental properties of elementary particles and cosmology'' N 0723-2020-0041, 
  and by RFBR Project 19-02-00346.


\end{document}